\documentclass[superscriptaddress,onecolumnpage,notitlepage,nofootinbib]{revtex4-1}

\pdfoutput=1
\usepackage{graphicx} 
\usepackage{tikz}
\usepackage{amsmath}
\usepackage{amssymb}
\usepackage{natbib}
\usepackage{color}
\usepackage{float}
\usepackage{epstopdf}
\usepackage[normalem]{ulem}
\usepackage{pgfplots,pgfplotstable}
\usepackage{soul}
\usepackage{accents}
\usepackage{nameref}
\usepackage{balance}
\usepackage[all]{nowidow}
\usepackage{wrapfig}
\usepackage{hyperref}

\usepackage{color}
\usepackage{gensymb}
\usepackage{amsmath}
\usepackage{dsfont}
\usepackage{xcolor}

\begin{document}

\title{Delayed buckling of spherical shells due to viscoelastic knockdown of the critical load}

\author{Lucia Stein-Montalvo}
\email{lsmontal@bu.edu}
\affiliation{
	Department of Mechanical Engineering, Boston University, Boston, MA, 02215.
}%

\author{Douglas P. Holmes}
\email{dpholmes@bu.edu}
\affiliation{
	Department of Mechanical Engineering, Boston University, Boston, MA, 02215.
}%

\author{Gwennou Coupier}
\email{gwennou.coupier@univ-grenoble-alpes.fr}
\affiliation{
	Universit\'e Grenoble Alpes, CNRS, LIPhy, F-38000 Grenoble, France.
}%

\begin{abstract}
\noindent We performed dynamic pressure buckling experiments on defect-seeded spherical shells made of a common silicone elastomer. Unlike in quasi-static experiments, shells buckled at ostensibly subcritical pressures (i.e. below the experimentally-determined load at which buckling occurs elastically), often following a significant time delay. While emphasizing the close connections to elastic shell buckling, we rely on viscoelasticity to explain our observations.
\noindent In particular, we demonstrate that the lower critical load may be determined from the material properties, which is rationalized by a simple analogy to elastic spherical shell buckling. We then introduce a model centered on empirical quantities to show that viscoelastic creep deformation lowers the critical load in the same predictable, quantifiable way that a growing defect would in an elastic shell. This allows us to capture how both the critical deflection and the delay time depend on the applied pressure, material properties, and defect geometry. These quantities are straightforward to measure in experiments. Thus, our work not only provides intuition for viscoelastic behavior from an elastic shell buckling perspective, but also offers an accessible pathway to introduce tunable, time-controlled actuation to existing mechanical actuators, e.g. pneumatic grippers.

\end{abstract}

\maketitle

\section{Introduction} 
\label{sect:intro}
Shell structures are lightweight and flexible. Largely owing to their curvature, they offer considerable strength with little material. As a result, shells are abundant in nature (e.g. eggshells and blood vessels) and design (e.g. fuel tanks and soda cans). However, slenderness also brings susceptibility to abrupt and often catastrophic deformations. Clearly, understanding how a thin, curved structure will lose stability -- and in particular at what load value this will occur -- is crucial.

We restrict our attention to spherical shells in the present work. The first-known quantitative prediction for the critical load $P_c$ in a perfect, spherical, elastic shell subjected to uniform pressure was produced  by Zoelly in 1915 \textit{via} linear eigenvalue analysis, and is given as:

\begin{equation} \label{ZoellyPc}
P_c = \frac{2 E}{\sqrt{3 (1-\nu^2)}} \eta^{-2}
\end{equation}
for a shell with Young's modulus $E$, Poisson's ratio $\nu$, and radius ($R$) to thickness ($h$) ratio $\eta \equiv R/h$.

Although this result is still widely accepted today, it severely overpredicts the buckling load observed in experiments. Recognizing this disrepancy, which is due to the extreme sensitivity to imperfections inherent to thin shells, scientists at the space agency NASA and collaborators introduced the \textit{knockdown factor} ($k_d$) in 1930. The quantity is defined as the ratio of the observed critical load $P_c^e$, to that predicted by the theory, i.e. $k_d \equiv P_c^e/P_c$. Based on surveyed experimental results~\cite{Lee2016,Wagner2020}, engineers at NASA settled for the extremely conservative design code of $k_d \approx 0.2$ for spherical shell structures~\cite{NASASP8032}.

Over the decades that followed, extensive work was dedicated to correcting the persistent overprediction of the critical pressure. This involved studies of the post-buckling behavior~\cite{VonKarman1939,Tsien1942}, and the imperfection sensitivity~\cite{Koiter1945,Koiter1969,Hutchinson1967,Bushnell1967,Krenzke1965,Koga1969} of thin spherical shells.
Yet, marked success arrived only recently, after Lee et al. developed a new fabrication technique for polymeric spherical caps~\cite{Lee2016a}. The authors used this method to create dominant dimple-like defects (larger than those naturally occurring in the shell) with systematic size variations~\cite{Lee2016}. These experiments, validated with finite element modeling (FEM) and numerical analysis, quantitatively showed for the first time how the imperfection depth lowers the critical load. 

Other contributions followed, including studies on spheres with similar dimple defects and sinusoidal equatorial undulations~\cite{Hutchinson2016}, large-amplitude dimples~\cite{Jimenez2017}, through-thickness defects~\cite{Yan2020} (a notable predecessor is Ref.~\cite{Paulose2013}), dent defects~\cite{Gerasimidis2020}, and probing force imperfections~\cite{Gerasimidis2018}, which collectively clarify the effect of the type of defect on the knockdown factor for spherical shells. 
More broadly, this long-awaited breakthrough provoked a new surge of progress in spherical shell theory (see e.g. Refs.~\cite{Hutchinson2016,Hutchinson2017,Audoly2020,HutchinsonEMLoverview}). These developments afford engineers the opportunity to design sturdy structures with more specific -- and permissive~\cite{Wagner2020} -- lower bounds on the load carrying capacity. This was the initial goal. In more recent years, though, a community of researchers has adopted, in a sense, the opposite goal: to design structures that buckle and snap on command~\cite{Holmes2019,Reis2015} for functions like colloidal self-assembly~\cite{Sacanna2010}, encapsulation~\cite{Shim2012}, inflatable snapping actuation~\cite{Keplinger2012,Gorissen2020}, and artificial muscle actuation~\cite{Yang2016}. 

The new understanding of defect sensitivity in a different light demonstrates the tunability of spherical shell buckling, and thus serves this ``buckliphilic"~\cite{Reis2015} community equally. In particular, the geometry and placement of a dominant defect prescribe, respectively, the buckling strength and the spot where an instability localizes. Recent extensions of this concept couple geometric defects with differentially swelling~\cite{Lee2019} or magneto-responsive~\cite{Yan2020b} materials to modify the knockdown factor over time. Relatedly, a more general study showed how a homogenous natural curvature -- which can be a proxy for nonmechanical stimuli like thermal expansion, changes in pH, or differential growth -- acts to raise or lower the knockdown factor in spherical shells~\cite{Holmes2020}.

Besides control over when instability occurs -- which depends on geometry, loading, and mechanics -- mechanical actuators generally rely on reversibility. Repeatable actuation calls for robust, elastic materials, and silicone rubbers like polydimethylsiloxane (PDMS) and vinyl polysiloxane (VPS) have answered this call in mechanics research~\cite{Reis2015b}. In addition to their elastic behavior, these elastomers are readily accessible and allow for fast, easy fabrication~\cite{Lee2016a}.

Recently, Djellouli, et al. combined these ingredients to produce a mechanical swimmer~\cite{Djellouli2017}. Quasi-static pressure cycles drove the device, a defect-seeded spherical shell made of the elastomer Dragon Skin\textsuperscript{TM} 30, to propel forward through a viscous fluid by buckling and unbuckling controllably. The authors propose that maintaining dimensionless quantities constant would allow for miniturization, with implications for drug delivery.  A natural extension of this work, and the motivation for the present study, is to seek control over the speed of swimming by adjusting the frequency and/or amplitude of pressure cycles. Largely because shell and fluid motion are highly coupled in swimming, and because of possible resonance with postbucking oscillations~\cite{Mokbel2021}, we expect this phase space to be complex. Thus, we set out to first isolate the shell buckling response, independent of fluid motion, to dynamic loading at pressures in the vicinity of the critical load.  

We fabricated imperfect spherical shells like those in Ref.~\cite{Djellouli2017}, and fixed them in place surrounded by air. A small nozzle allows for internal pressure control, through which we step-load the shells -- that is, we abruptly apply, and then maintain, a pressure load. Explicitly, we reduce the pressure inside the shell cavity, creating a negative inside-outside differential pressure. For simplicity, we will refer to this pressure difference in terms of its magnitude.
These straightforward experiments produced surprisingly rich results. Even for loads below the experimentally measured elastic critical pressure (which we loosely call ``subcritical" herein), we consistently observe buckling. Further, this buckling at subcritical loads occurs abruptly, often after an extended period of very slow deformation perhaps mistakeable for stability. We observe singular thresholds and a delay time which increases monotonically as pressure decreases -- this contrasts the findings of a recent numerical study on dynamic step loading of spherical shells which are much thinner than our own~\cite{Sieber2019}.

As discussed, geometric imperfections can lead to buckling at lower-than-expected loads. In this case, though, the shell defects are already accounted for. Devoid of any plausible geometric explanation for this strange buckling behavior, our results require a closer examination of the materials. Although silicone elastomers are selected precisely because they behave elastically in most settings, they are in fact prone to time-dependent molecular rearrangement, and hence are viscoelastic~\cite{Lin2009, Yuk2017, Urbach2020}. Thus, they behave differently depending on how fast they are loaded, and exhibit both stress-relaxation (softening when subjected to a constant strain) and creep (deformation over time under constant stress).

In the present work, we rely on this viscoelasticity to account for our observations, while emphasizing the close connections to elastic spherical shell buckling. The structure is as follows: First, we introduce key aspects of our experiments in Sect. \ref{sect:matexps}. Specifically, we report the viscoelastic material properties, describe the geometry of our imperfect spherical shells, and briefly introduce our experimental setup. In Sect.~\ref{sect:3regimes}, we present an overview of our findings. We address the critical pressure conundrum in Sect. \ref{sect:press} by defining two pressure thresholds: the elastic critical pressure, and the lower viscoelastic critical pressure, which can be related through the limiting material properties. These thresholds separate three regimes: immediate buckling, delayed buckling, and no buckling (stable).
In Sect. \ref{sect:defect}, we introduce an analogy wherein viscoelastic creep deformation lowers the critical load in the same way that a growing dimple-like defect would in an elastic shell. This provides insight about the pre-buckling deformation (Sect.\ref{sect:wc}), and reveals how the delay time preceding buckling depends on the imposed pressure, shell geometry and material properties (Sect.\ref{sect:tc}). Finally, we offer concluding remarks in Sect. \ref{sect:concl}.

\section{Materials, geometry and methods} \label{sect:matexps}
We have performed dynamic, step-loaded pressure buckling experiments on soft, viscoelastic spherical shells. Here, we briefly summarize the material properties, shell fabrication and geometry, and dynamic loading methods.

\subsection{Material characterization}\label{sect:materialcharacterization}
The shells used in our experiments are made of the elastomer Dragon Skin\textsuperscript{TM} 30 (manufactured by Smooth-On; Poisson's ratio $\nu = 0.5$~\cite{Djellouli2017}). We assume that our viscoelastic material can be described by the Standard Linear Solid (SLS) model, the simplest linear model that captures both stress-relaxation (the decreasing stress response over time for a structure subjected to a constant strain) and creep (deformation under a prolonged constant stress)~\cite{Zener1948}. The SLS model describes limited creep behavior, i.e. creep deformation does not progress indefinitely, nor does the modulus eventually go to zero. 

According to the SLS model, the modulus relaxes over time according to:
\begin{equation}\label{relaxmod}
E(t) \equiv \frac{\sigma(t)}{\varepsilon}  = E_{\infty} + E_1 e^{-t/\tau_{\sigma}}
\end{equation}
where $\sigma(t)$ is the time-varying stress, $\varepsilon$ the constant strain, and $\tau_{\sigma}$ the relaxation time. The parameter $E_1 \equiv E_0-E_{\infty}$ quantifies the total stiffness lost as the elastic modulus $E_0$ decreases to the long-term (equilibrium) modulus $E_{\infty}$, where $E_0 \geq E_{\infty}$. The function \eqref{relaxmod} is known as the relaxation modulus. It is related by Laplace transform to the creep compliance function, which describes the temporally increasing strain $\varepsilon(t)$ of the SLS element under imposed constant stress $\sigma$~\cite{Lakes2009}: 

\begin{equation}\label{creepcomp}
J(t) \equiv \frac{\varepsilon(t)}{\sigma} = J_0 + J_1 \big( 1-e^{-t/\tau_{\varepsilon}} \big)
\end{equation}
where $J_0 = E_0^{-1}$, $J_1 \equiv J_{\infty}-J_0$ with $J_{\infty} = E_{\infty}^{-1} \geq J_0$, and $\tau_{\varepsilon}$ is the retardation time.

We performed uniaxial tension and stress-relaxation tests using the tensile testing machine Instron 5943 to identify the parameters in Eq. \eqref{relaxmod} (details are provided in Sect. \ref{sect:creeptests}.) We found the relaxation time to be $\tau_{\sigma} = 0.78 \pm 0.49$ s. As for the moduli, we determined $E_0 = 0.59 \pm 0.04$ MPa and $E_{\infty} = 0.54 \pm 0.08$ MPa. Reported errors throughout the text correspond to one standard deviation unless otherwise noted\footnote{For quantities with few measurements, the approximate standard deviation is reported as one-fourth of the error range.}. The resulting ratio of the mean long-term modulus to the mean instantaneous one is $\bar{E} \equiv E_{\infty}/E_0 = 0.91$, which is central to the analysis beginning in Sect. \ref{sect:press}. 

Since our materials have a relatively low relaxation strength, defined as $\Delta = E_1/E_{\infty}$~\cite{Lakes2009}, Eq. \eqref{relaxmod} and the inverse of Eq. \eqref{creepcomp} differ negligibly, i.e. $J^{-1}(t) \approx E(t)$ (see Sect.~\ref{sect:creeptests}, Fig.~\ref{fig:creepparams}). While creep is the relevant process in our experiments, our primary aim is to draw connections to elastic shell theory, which relies on the modulus $E$. Thus, we will use this convenient fact to interchange the representation of these two mechanisms in our analysis. 

\subsection{Shell geometry}\label{sect:geom}
Spherical shells were fabricated following the bi-molding method from Ref.~\cite{Djellouli2017}, which is detailed in Sect. \ref{sect:shellfab}. The shells, made of two hemispheres seamed with a thin layer of diluted polymer, all have an outer radius of $R_o = R+h/2 = 25$ mm. The thickness $h \in [1$ mm, $5$ mm$]$, such that $\eta \in \{24.5,12.0, 6.6,4.5\}$. 

\begin{figure}[h!] 
	\centering
	\includegraphics[width=0.3\linewidth]{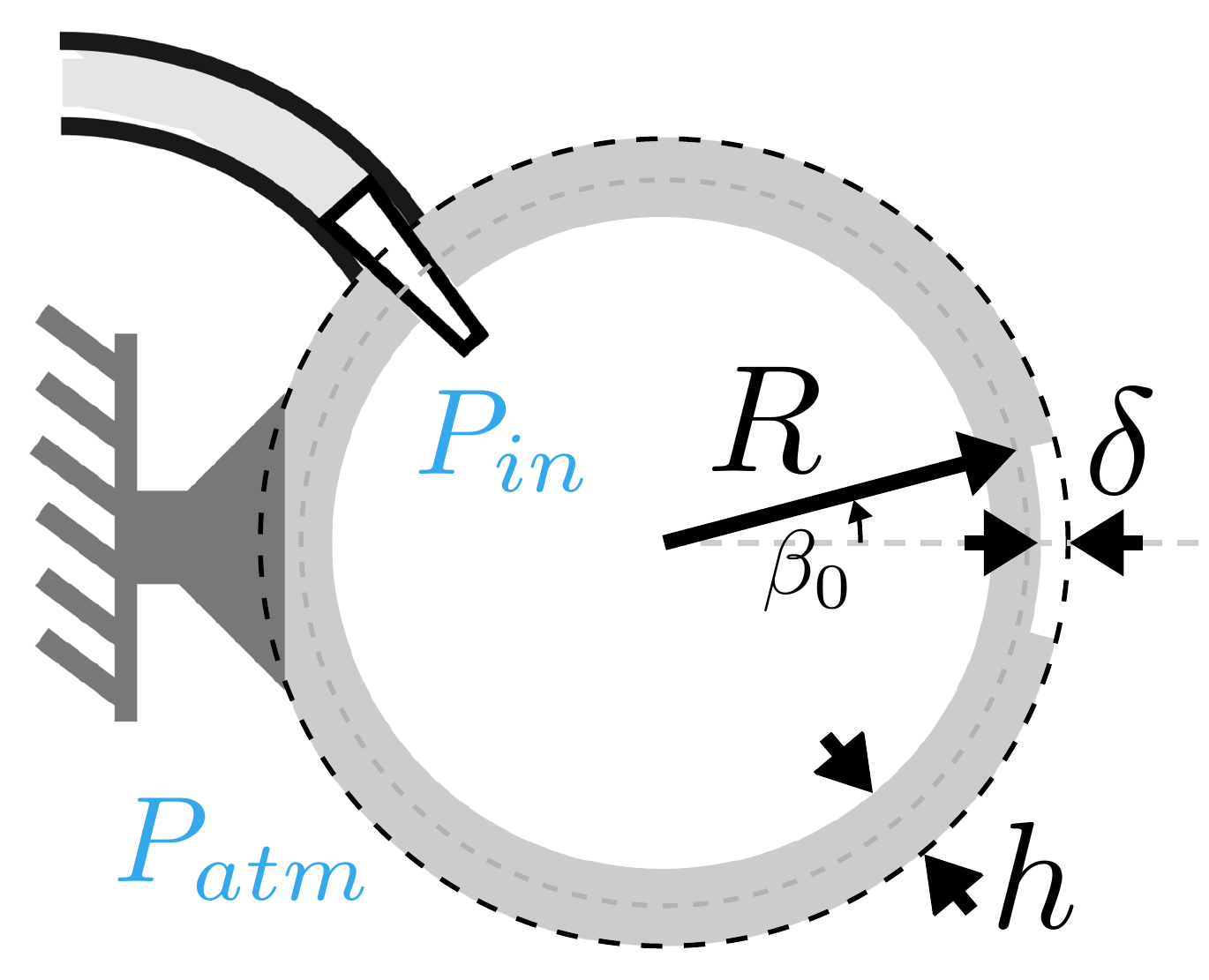}
	\caption{Schematic of the clamped shell with an inlet for pressure control. The flexible tube connects at its other end to a vacuum tank so that when the line is open, $P_{\text{in}} < P_{\text{atm}}$ (or $P \equiv |P_{\text{in}} - P_{\text{atm}}| > 0$). The relevant parameters are labeled: the nominal thickness $h$ and midline radius $R$, imperfection depth $\delta$, and half-angular width of the imperfection $\beta_0$.}
	\label{fig:shellparams}
\end{figure}

To control the location and direction of buckling, each shell is seeded with an imperfection, where the thickness is reduced by an amount $\delta\in [0.40,0.81]$ mm, such that $\bar{\delta} \equiv \delta/h \in [0.08,0.76]$ in a circular region spanning a half-angle of $\beta_0 \approx \pi/24$ radians. The  shell parameters are shown schematically in Fig. \ref{fig:shellparams}, and the effect of this defect is discussed further in Sect. \ref{sect:PceSect}.

Following e.g. Ref.~\cite{Lee2016}, we also introduce the parameter $\lambda$, defined as
\begin{equation}\label{lambda}
\lambda = \big(12 (1-\nu^2)\big)^{1/4} \eta^{1/2} \beta_0,
\end{equation}
which describes the defect geometry in the context of spherical caps~\cite{Kaplan1954,Taffetani2018}. For our shells in order of increasing thickness, $\lambda \in  \{1.038, 0.716, 0.612, 0.492\}$. 

\subsection{Step loading}\label{sect:steploading}
A flexible tube was connected on one end to the inside of the shell, and on the other to a vacuum tank. An electrovalve interrupting this channel allowed us to abruptly remove air from the inner volume of the shell, creating a pressure difference of magnitude $P \in [0.3$ kPa, $46.5$ kPa$]$ (see Fig. \ref{fig:shellparams}), which we monitor with a pressure sensor. The pressure load is maintained for either the time it takes the shell to buckle, or $t_{hold} \in [5 s,360 s]$. For details on the experimental setup, see Sect. \ref{sect:pressurecontrol}.

\section{Three  regimes}\label{sect:3regimes}
\begin{figure}[!h] 
	\centering
	\includegraphics[width=1\linewidth]{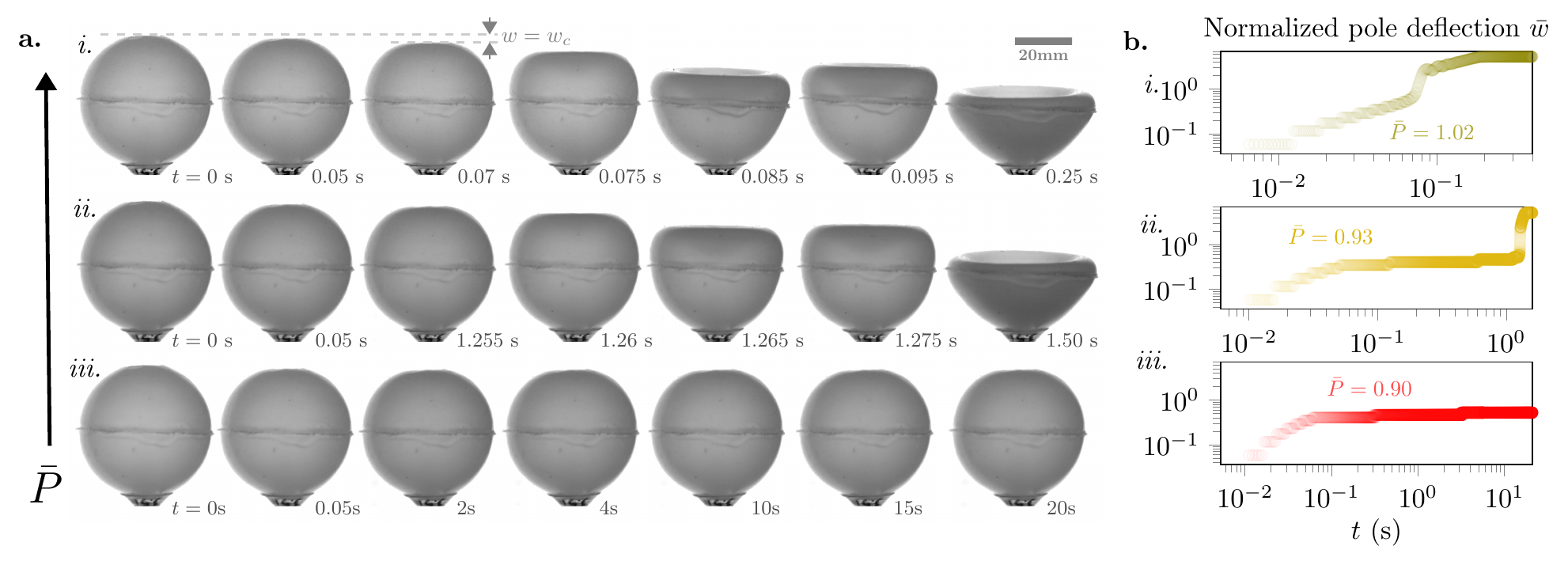}
	\caption{(a.) Selected high-speed camera images and (b.) corresponding plots of the pole deflection normalized by the shell thickness, i.e. $\bar{w} = w/h$, show the typical response of a shell ($\eta= 6.6$) to step loading at relatively \textit{i.} high (supercritical), \textit{ii.} moderate (subcritical), and \textit{iii.} low pressures. The pole deflection $w$ is measured from the initial state,  marked in a,\textit{i.} by the upper dashed gray line. 
		At or above the elastic buckling pressure (e.g. $\bar{P} \equiv P/P_c^e = 1.02$), the shell quickly buckles ($t_c \approx 0.07$ s). The corresponding critical pole deflection $w_c$ is indicated in a,\textit{i.}
		Even at subcritical pressures, e.g. $\bar{P} = 0.93$, the shell eventually buckles ($t_c \approx 1.26$ s). This collapse follows a deceleration in the pole deformation at $t \approx 0.05$ s, and a subsequent period of slow, constant-rate deformation.
		For pressures below a second threshold, e.g. $\bar{P} = 0.90$, the shell does not buckle. 
	}
	\label{fig:seqpole}
\end{figure}

For each of our shells, the immediate response to any non-negligible pressure load was qualitatively the same: The shell compresses as soon as the pressure is felt, and deformation quickly localizes at the unclamped pole (in the vicinity of the imperfection), forming a dimple-like depression with a deflection depth $w$ (see Fig. \ref{fig:seqpole} a.) Beyond this early behavior, which occurs in approximately the first 0.05 seconds, three regimes were evident from our experiments.

If the imposed pressure is high, the initial fast rate of pole deformation is maintained, and the shell quickly buckles -- that is, the pole inverts, driving global collapse (Fig. \ref{fig:seqpole}, \textit{i}). In postbuckling, which we do not study in detail here, the pole region is completely inverted and oscillations occur before stability is reached again when $w \approx 2R$. We refer to this regime, wherein the shell behaves elastically throughout deformation, as the {\em immediate buckling regime}. The lowest pressure at which we observe this buckling behavior defines the experimental elastic critical load $P_c^e$. Due to the seeded defects and the non-negligible thickness of our shells, the experimentally-determined value $P_c^e$ differs from the theoretical critical pressure $P_c$ (Eq.~\eqref{ZoellyPc}) for thin, perfect elastic shells. For details, see Sect. ~\ref{sect:PceSect}.

At slightly lower pressures, the deformation rate slows considerably following the fast response to loading, at a transition time ($t \approx 0.05$ s). The dimple slowly deepens (i.e. $w$ increases; See Fig. \ref{fig:seqpole}, \textit{ii}), before an abrupt acceleration after some time $t_c \in [0.09$ s, $17.09$ s$]$ signifies buckling. We define this intermediate regime as the {\em delayed buckling regime}. At still-lower pressures, slow pole motion eventually stops, and the shell settles into indefinite stability for as long as the load is maintained. We call this third regime the {\em stable regime} (Fig. \ref{fig:seqpole}, \textit{iii}).

The value of the pressure which separates the delayed buckling and stable regimes, and hence marks the boundary of whether collapse will occur, is clearly of interest. This lower pressure threshold was more or less constant for all of our shells when normalized by the elastic load. In other words, the reduced critical pressure is independent of geometry. With this nudge toward the materials, we proceed to rationalize these findings.

\section{Pressure thresholds via modulus ratio}\label{sect:press} 
Since we know our materials are viscoelastic, we can presume that the slow pole deformation under constant pressure is an exhibition of creep. This would situate our observations in the terrain of \textit{creep buckling}. Creep buckling was introduced in the literature in 1951 \cite{Rosenthal1951}. The bulk of the work in this field was developed in the thirty or so years that followed, and was aimed at understanding creep collapse that occured on timescales of hours or even days in metallic, mono-resin materials, and reinforced concrete. However, general theories emerged, which are illuminating when applied to our elastomer shells. In particular, Hayman~\cite{Hayman1978,Hayman1981} and others~\cite{Hoff1973,Minahen1993} proposed that a viscoelastic structure that buckles due to creep may be treated as an equivalent elastic structure with a lower critical load. The main result is that the lower threshold -- which we will henceforth refer to as the viscoelastic critical pressure $P_c^v$ -- is directly related to the long-term modulus. For spherical shells, the lower critical pressure $P_c^v$ may be found by simply replacing Young's elastic modulus $E$ with $E_{\infty}$ in calculating the elastic critical load (Eq.~\eqref{ZoellyPc}). Denoting normalization by the experimentally measured elastic critical pressure $P_c^e$ with an overbar, i.e. $\bar{P} \equiv P/P_c^e$ and $\bar{P}_c^v \equiv P_c^v/P_c^e$, that is: 

\begin{figure}[h!]
	\centering
	\includegraphics[width=0.5\linewidth]{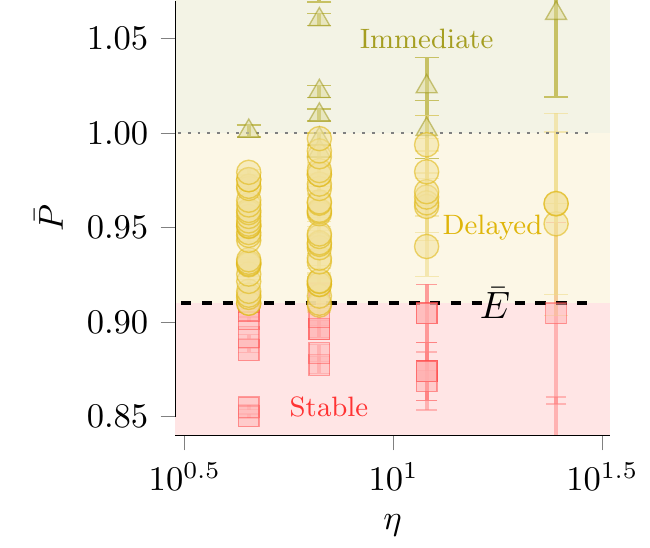}
	\caption{
		Phase plot depicting the three regimes, separated by two pressure thresholds. The elastic critical pressure (thin dotted line, $\bar{P}=1$) was experimentally determined as the minimum pressure at which a slowdown of dynamics does not preceed buckling. We expect ``immediate" (elastic) buckling for an imposed pressure $\bar{P} \geq 1$, which corresponds to the green region. The theoretical viscoelastic critical pressure $\bar{P}_c^v = \bar{E}$ (thick dashed line) is determined from \eqref{ratio} using the material parameters measured independently in stress-relaxation tests (see Sect. \ref{sect:materialcharacterization}). Between $\bar{P}=1$ and $\bar{P}_c^v$ (yellow region), viscoelastic creep can explain delayed instability at subcritical pressures. Below $\bar{P}=\bar{E}$ (red region), limited creep for our material is insufficient to cause buckling, so we expect indefinite stability. Green triangles, yellow circles, and blue squares represent, respectively, experiments which buckled elastically, buckled after a time delay, and did not buckle. Error bars represent one standard deviation.
	}
	\label{fig:phase}
\end{figure}

\begin{equation}\label{ratio}
\bar{P}_c^v = \bar{E}.
\end{equation}

In Fig.~\ref{fig:phase}, we show that \eqref{ratio} agrees with our experimental data very well, solidifying the notion that viscoelasticity is indeed the cause for the subcritical buckling we observe. The dashed line marks the theoretical lower limit where creep bucking may be observed, $\bar{P}_c^v = \bar{E}$, which for our materials (see Sect.\ref{sect:materialcharacterization} \& Sect.~\ref{sect:creeptests}) is $0.91$. 

Explicitly, Eq.~\eqref{ratio} ignores the actual mechanism that leads to instability, creep deformation, in favor of a straightforward way to determine the minimum buckling pressure for creep-limited materials, which is perhaps the most crucial information for any design goal. Given the elastic (viscoelastic) critical pressure and the experimentally-determined instantaneous and long-term moduli, the ratio in Eq.~\eqref{ratio} readily predicts the viscoelastic (elastic) critical pressure. 

This view is effective and completely general with respect to geometry and material properties. However, it provides no information about deformation or the time delay that preceeds buckling. Besides that these features are of fundamental interest, understanding this delay mechanism will offer a route to including controllable delays in elastomer device design. We address these open questions in the following section.

\section{Creep deformation as an evolving defect}\label{sect:defect}
As we have seen, the efficient, modulus-based approach in Sect.~\ref{sect:press} connects the limiting critical pressure of a viscoelastic shell to that of the equivalent elastic shell. It leaves questions, however, about the time it takes a subcritically-loaded shell to buckle, and the underlying pre-buckling deformation. Traditional analytical approaches to capture the critical time and/or deflection for creep buckling involve incorporating calculated quantities for stress and strain into the constitutive model (in our case Eq.~\ref{creepcomp}). Instability may be identified by solving the eigenvalue problem of the governing differential equations, or by the quasi-static ``critical strain approach"~\cite{Gerard1958} wherein the critical strain must be known or assumed \textit{a priori}, and the corresponding time is directly solved for~\cite{Miyazaki2015}. These methods require precise representations of the stresses and strains throughout deformation, and have met moderate success in capturing experimental behavior for simple structures like columns~\cite{Hoff1954,Hoff1956}, trusses and arches~\cite{Huang1967}, plates and even cylinders~\cite{Gerard1958}. (See Ref.~\cite{Miyazaki2015} for a review of the relatively recent work on creep buckling of shell structures, or Ref.~\cite{Hoff1973} for an earlier review on creep buckling of plates and shells.)

A clear problem with these approaches is that it is generally assumed that a shell undergoing creep will lose stability at the same strain as its elastic counterpart~\cite{Gerard1956,Hoff1975}. However, it has been noted that this assumption often leads to underprediction of the critical displacement and time~\cite{Obrecht1977}. Indeed, although the immediate and delayed buckling regimes appear qualitatively very similar in terms of deformation in our experiments (see Fig. ~\ref{fig:seqpole}, \textit{i. \& ii.}), we observe that shells which creep for longer sustain more deformation before buckling. 

Complex geometries like imperfect spherical shells and their nonlinear deformations introduce significant analytical difficulties of their own. Creep buckling in spherical caps and complete spherical shells has primarily been studied with numerical analyses~\cite{Huang1965, Shi1970, Jones1976, Miyazaki1977, Xirouchakis1980}, which do not produce closed-form solutions for when instability occurs. Few experiments exist for comparison to these results, and attempts to replicate limited experimental creep buckling behavior for spherical shells have largely been unsuccessful~\cite{Shi1970,Leckie1976}.

\begin{figure}[h!]
	\centering
	\includegraphics[width=0.8\linewidth]{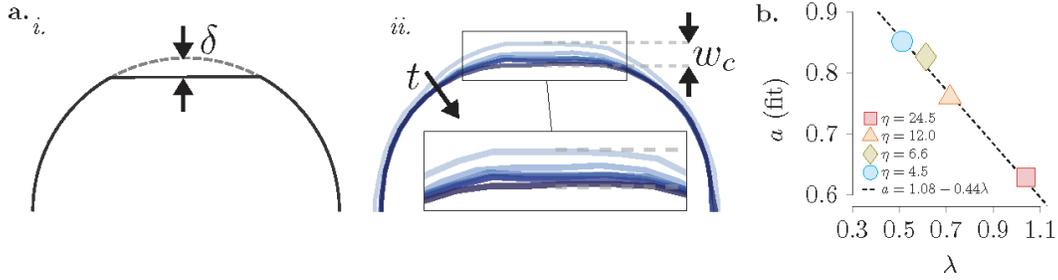}
	\caption{Pre-buckling pole deformation ($w \leq w_c$) is qualitatively similar to a geometric defect of increasing depth $\delta$. (a.) \textit{i.} Schematic of dimple-like defect in an elastic spherical shell, after e.g. Ref.~\cite{Lee2016}. \textit{ii.} Edge contours from high-speed images at $t = 0, 0.07, 0.14, 0.21, 0.28$ s and $t_c = 0.35$ s (darkening blue corresponds to increasing time) for $\eta = 4.5$ and $\bar{P} \approx 0.97$. After initial compression, deformation localizes to the pole and progresses. (b.) Our fitting parameter $a \approx f(\lambda) = 1.08-0.44 \lambda$.  
	}
	\label{fig:defect_lambda}
\end{figure}

A more enlightening approach relies on the observation that the pre-buckling deformation approximately amplifies the initial defect (see Fig.~\ref{fig:defect_lambda}a). This suggests that we may be able to draw an analogy between creep deformation and a growing imperfection in an elastic shell.  This concept was proposed by Hayman in 1981, but the author conjectured that while his ``locus of critical points" approach offered intuition, it lacked predictive power for all but simple, statically determinate structures~\cite{Hayman1981}. To our knowledge, the approach has not been implemented besides in the original work, when it was validated against a small number of experiments on concrete three-pin arches. 

One key challenge was that the effect of the defect must be quantifiable. Indeed, as discussed in Sect.~\ref{sect:intro}, the imperfection sensitivity of shell structures has only recently become experimentally tractable. In what follows, we rely on the work of Lee and collaborators~\cite{Lee2016}, and center our analysis upon quantities that are readily determined through experiments, to derive a practical and predictive model for creep buckling in our shells.

Knowing that an imperfection ``knocks down" the critical load in an elastic structure, in this view quasi-static creep deformation has the same effect. In other words, creep deformation behaves like an evolving imperfection by progressively knocking down the critical load. It follows that creep collapse occurs when the critical load associated with the ``imperfect" creep-deformed structure falls to the value of the applied load. Conversely, the imperfection size associated with a given applied load should correspond to the critical deformation at which creep buckling occurs. It is important to note that this was buried but implied in the approach of Sect.~\ref{sect:press}; the two arguments are complementary.

To make this analogy quantitative, we turn to the recent literature on geometric imperfections in elastic spherical shells. The true defects in our shells are characterized by a local reduction in thickness as in Ref. \cite{Yan2020} (see Sect. \ref{sect:PceSect}). The pole deflection generated during creep, however, is qualitatively more similar to a dimple-like imperfection where the thickness does not change, but the curvature of the shell midline does. Because the midline curvature of our shells is unaffected by the thickness reduction except at the discontinuity at either edge of the defect profile, we consider our shells initially ``perfect" in the dimple sense. For the present analysis, we rely on the findings of Lee et al.~\cite{Lee2016}, which are in agreement with Refs.~\cite{Hutchinson1967, Hutchinson2016, Jimenez2017}.

The key finding of their work, for our purposes, is that the knockdown factor for a given shell $k_d = P_c^e/P_c$ is a function of $\bar{\delta} = \delta/h$ which initially decreases for increasing $\bar{\delta}$, then reaches a plateau. The authors present an empirically-determined function describing the lower bounding envelope over the range of $\lambda$ they study, which takes the form $k_d = a + b/(c+\bar{\delta}$). We find that this functional form describes their individual curves sufficiently well. 

Primarily because $\lambda$ for our thick shells is below the range studied in Ref.~\cite{Lee2016}, we cannot directly extract the results relevant to our work. Instead, we expect that these general trends will hold. We assume that we can simply replace the imperfection depth $\delta$ with the pole deflection $w$. Because we only consider the additional knockdown due to creep deformation, and not that due to the inital defect, we take the reference pressure as the experimental critical pressure $P_c^e$ that corresponds to the initially imperfect shell. Then we define a general viscoelastic knockdown function
\begin{equation}\label{kdv_wform}
k_d^{v}(w) = \bar{P}_c(\bar{w}) = a + \frac{b}{c+\bar{w}},
\end{equation}
where $\bar{w}=w/h$, and $a$, $b$, and $c$ are yet unknown.

Since $w$ increases according to the creep strain rate, an alternative form of Eq.~\eqref{kdv_wform} specifies the time dependence. Approximating the magnitude of the circumferential strain to first order as $\varepsilon \approx w/R$, we can say $\bar{w} \approx \eta \varepsilon(t)$. From Eq.~\eqref{creepcomp} this means $\bar{w} \approx \eta \sigma J(t)$, which is approximately $\eta \sigma/E(t)$ since the creep compliance function $J(t)$ and the inverse of the relaxation modulus $1/E(t)$ are nearly indistinguishable for our material. At early times the shell behaves elastically, so we assume Hooke's Law applies, \textit{i.e.} $\sigma \approx E_0 \epsilon(t_t)$ at a transition time $t_t$ between the elastic and creep stages of deformation. Further, we assume $\epsilon(t_t) \approx \epsilon_c^e$, where $\epsilon_c^e \approx w_c^e/R$ is the critical strain corresponding to $\bar{P} = 1$, when the shell buckles immediately following elastic deformation. Then by Eq.~\eqref{relaxmod}, the normalized pole deflection increases with time following the relation
\begin{equation}\label{wbarstrainform}
\bar{w}(t) \approx \frac{E_0 \: \bar{w}_c^e}{E_{\infty}+E_1 e^{-t/\tau_{\sigma}}}.
\end{equation} 
Note that due to our simplified representations of stresses and strains in Eq.~\eqref{wbarstrainform}, all relations that follow are approximations, despite that we present them as equalities for simplicity. Substituting Eq.~\eqref{wbarstrainform} in Eq.~\eqref{kdv_wform}, gives a time-dependent version of the generalized viscoelastic knockdown function:
\begin{equation}\label{kdv_tform}
k_d^{v}(t) = \bar{P}_c(t) = a + \frac{b}{c+\frac{E_0 \: \bar{w}_c^e}{E_{\infty}+E_1 e^{-t/\tau_{\sigma}}}}.
\end{equation} 

It remains to determine the three unknown quantities, which should explain how sensitive the critical pressure is to deformation (Eq.~\eqref{kdv_wform}) and how quickly the critical pressure decreases (Eq.~\eqref{kdv_tform}) for each shell. To do so, we constrain the functions, enforcing what we know about the limiting behavior. The critical deflection required for buckling at the elastic limit where $\bar{P}=1$ is an experimentally-determined value, $w_c^e$, which differs for each shell based on geometry. Since for buckling to occur, $P_c(w) = P$, from Eq. \eqref{kdv_wform} this condition is stated as:
\begin{equation}\label{cond1}
1 = a + \frac{b}{c+\bar{w}_c^e}.
\end{equation}

We also know from Eq.~\eqref{ratio} that buckling will not occur below $k_d = \bar{P}_c^v = \bar{E}$. Taking the limit of the right hand side of Eq.~\eqref{kdv_tform} as $t$ approaches infinity gives a second constraint: 
\begin{equation}\label{cond2}
\bar{E} = a + \frac{b}{c+\frac{\bar{w}_c^e}{\bar{E}}}.
\end{equation}

Solving Eqs. \eqref{cond1} \& \eqref{cond2} simultaneously gives

\begin{equation}\label{b}
b = \frac{(a-1)(a -\bar{E}) \bar{w}_c^e}{\bar{E}}
\end{equation}
and 
\begin{equation}\label{c}
c = \frac{-a \bar{w}_c^e}{\bar{E}},
\end{equation}
which we insert into Eq.~\eqref{kdv_wform} to arrive at:
\begin{equation}\label{kdv_w_full}
\bar{P}_c(w) = a + \frac{(1-a) (\bar{E}-a) \bar{w}_c^e}{\bar{E} \bar{w}- a \bar{w}_c^e}
\end{equation} 
which describes the critical pressure for a given degree of pole deflection. If $\bar{w}$ is large enough that $\bar{P}_c$ is lowered to the imposed dimensionless pressure $\bar{P}$, in theory buckling will occur.

Similarly, inserting the expressions for $b$ and $c$ in Eq.~\eqref{kdv_tform} gives
\begin{equation}\label{kdv_t_full}
\bar{P}_c(t) = a + (a-1)(a-\bar{E})\bigg(\frac{E_0 \bar{E}}{E_{\infty} + E_1 e^{-\frac{t}{\tau_{\sigma}}}}-a \bigg)^{-1}
\end{equation}
which specifies the time-dependence, according to the SLS model, of the knockdown to the critical pressure that occurs as the pole deflection progresses. Again, if $t$ is such that $\bar{P}_c = \bar{P}$, we expect collapse to occur.

Note that from Eq.~\eqref{kdv_t_full}, it is clear that any explicit dependence on geometry is contained in $a$. When $a$ is left as a fitting parameter for the curves defined by Eq.~\eqref{tc} (see Fig.~\ref{fig:tc}a), we find that it is a decreasing function of the geometric defect parameter $\lambda$, which is in general agreement\footnote{We fit a function of the form $k_d = a + b/(c+ \bar{\delta})$ to the numerical data over $\lambda \in [1,5]$ in Fig. 6b of Ref.~\cite{Lee2016}, and found $a \approx 0.5 - 0.14 \lambda$.} with Ref.~\cite{Lee2016}. In particular, we determine $a \approx 1.08-0.44 \lambda$ (Fig. \ref{fig:defect_lambda}b) for our shells. We refer to this linear function henceforth as $f(\lambda)$, and substitute it accordingly for $a$ to emphasize the deduced connection between our fitting parameter and the defect geometry. 

The deflection (Eq. \eqref{kdv_w_full}) and time (Eq. \eqref{kdv_t_full}) forms of the viscoelastic knockdown relations lead to expectations for how the critical displacement and critical time should depend on the material properties, geometry and the imposed pressure. We assess both in the following subsections.

\subsection{Critical deflection}\label{sect:wc}
\begin{figure}[h!]
	\centering
	\includegraphics[width=0.5\linewidth]{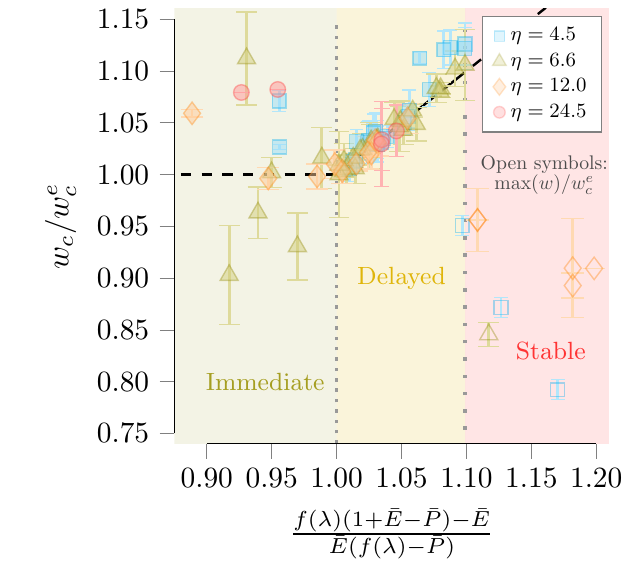}
	\caption{The critical pole deflection $w_c$ depends on $P$, the material properties, and the geometry through $a = f(\lambda)$ and $P_c^e$: when creep occurs (delayed buckling, yellow region, to the left of the dotted line at $1$ on the horizontal axis which marks $\bar{P} = 1$), the shell must deform more than $w_c^e = w_c(P_c^e)$ to ``knock down" the critical pressure until it coincides with the imposed pressure $P$, and stability is lost. This is represented by the diagonal line (Eq.~\eqref{wc}), which matches our data well until the lower pressure limit for buckling is approached. This asymptote is marked by the loosely dotted line with abscissa $1/\bar{E}$, which corresponds to $\bar{P} = \bar{E} = P_c^v/P_c^e$ (Eq. \eqref{ratio}). Open symbols in the red (stable) region represent the maximum (equilibrium) $w/w_c^e$ for experiments that did not buckle. As expected, these points fall below the theoretical line, which projects the deflection necessary for buckling. Our model does not extend to the immediate buckling regime (green region). Error bars are approximately one standard deviation. 
	}
	\label{fig:wc}
\end{figure}
Evaluating Eq. \eqref{kdv_w_full} at $P=P_c(\bar{w}_c)$, replacing $a$ with $f(\lambda)$, and solving for $\bar{w}_c/\bar{w}_c^e$ gives the following expression for the dimensionless critical deflection:  
\begin{equation}\label{wc}
\frac{w_c}{w_c^e} = \frac{f(\lambda) (1+\bar{E}-\bar{P})-\bar{E}}{\bar{E} (f(\lambda)-\bar{P})}
\end{equation}
which is valid when creep occurs ($\bar{P}_c^v = \bar{E} < \bar{P} < 1$, where the first equality refers to Eq. \eqref{ratio}) and tells us the pole deflection we should expect at (or, that is required for) buckling. Eq. \eqref{wc} is plotted against our data in Fig.~\ref{fig:wc}, showing good agreement until $\bar{P} \approx \bar{P}_c^v$. Near this asymptote, marked by the loosely dotted line, the critical deflection observed in experiments exceeds the predicted value. We discuss this deviation further in Sect.~\ref{sect:tc}.

The results in Fig.~\ref{fig:wc} are consistent with the trend observed in the creep buckling literature, where it is suggested that a shell undergoing creep prior to buckling sustains more deformation before collapse than its counterpart that buckles elastically. The explanation is that as the imposed pressure decreases, more deformation is required to sufficiently ``knock down" the critical pressure for buckling to occur. 

Creep does not occur before buckling when $\bar{P} \geq 1$ (the green region in Fig.~\ref{fig:wc}), where the shell behaves elastically. In this region, where Eq.~\eqref{wc} does not apply, we do not observe a clear trend in the critical deflection. Further, for experimental points that do not buckle, the deformation was insufficient to reduce the critical pressure to the value of the relatively low applied load. Accordingly, the maximum pole deflection (open symbols in Fig. \ref{fig:wc}) falls below theoretical curves when $\bar{P} < \bar{P}_c^v$ (red region in Fig. \ref{fig:wc}). Taking the limit as $t$ approaches infinity in either Eq.~\eqref{creepcomp} or Eq.~\eqref{relaxmod} gives $\varepsilon_{\text{max}} = \sigma/E_{\infty}$, so $w_{\text{max}} \approx R \sigma/E_{\infty}$. Meanwhile, Hooke's law provides an estimate for the critical deflection if $\bar{P} = 1$ is imposed: $w_c^e = R \sigma_e/E_0$. Then in theory, if $\bar{P} < \bar{P}_c^v$, $w_{\text{max}}/w_c^e \approx \sigma/(\bar{E} \sigma_e) \sim \bar{P}/\bar{E}$. We do not attempt to verify this scaling with our sparse data in this region, besides noting that as $\bar{P}$ approaches $\bar{P}_c^v$, we would expect $w_{\text{max}}/w_c^e$ to approach 1 (from below). Our data appears to support this conjecture. 

\subsection{Critical time}\label{sect:tc}
We have seen that the pole deflection is analogous to a dimple-like defect. Incorporating the viscoelastic material model, in turn, tells us how the critical pressure is expected to decrease over time. This offers a means to explain the critical buckling time, which increases monotonically for decreasing pressure. To do so, we evaluate Eq.~\eqref{kdv_t_full} at $\bar{P}=\bar{P}_c(t_c)$ and solve for the dimensionless critical time $t_c/\tau_{\sigma}$:
\begin{equation}\label{tc}
\frac{t_c}{\tau_{\sigma}} = \ln \bigg( \frac{(\bar{E}-1) (f(\lambda) (\bar{E}-\bar{P}+1)-\bar{E})}{\bar{E} (f(\lambda)-1) (\bar{E}-\bar{P})} \bigg).
\end{equation} 

\begin{figure}[h!]
	\centering
	\includegraphics[width=0.9\linewidth]{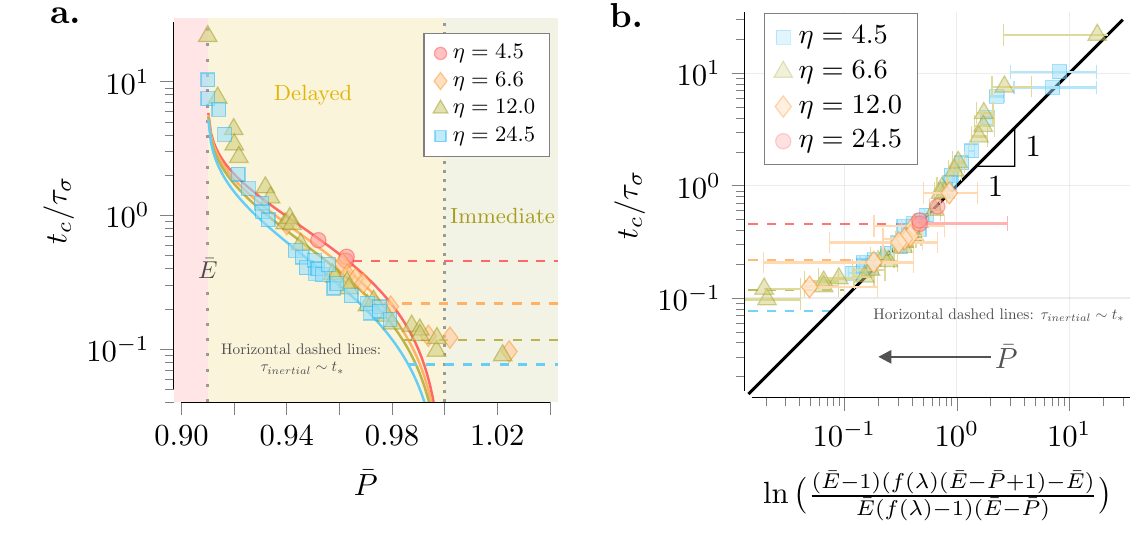}
	\caption{As the dimensionless applied pressure increases from $\bar{P}_c^v = 0.91$ to the elastic critical pressure $\bar{P} = 1$, the buckling time decreases monotonically according to Eq.~\eqref{tc}, which specifies how long it takes for the pole to deform enough to decrease the critical pressure to the applied one.
		(a.) Experimental data (markers) and color-corresponding curves from Eq.~\eqref{tc}, which were fit to obtain $f(\lambda) = 0.62, 0.76, 0.81, 0.85$ (plotted in Fig.~\ref{fig:defect_lambda}b). for $\eta = 4.5, 6.6, 12.0, 24.5$, respectively. The densely dotted vertical line marks $\bar{P} = 1$. Near this limit, the inertial (elastic) timescale, indicated by horizontal dashed lines, sets $t_c$ for each shell. The loosely dotted vertical line marks the asymptotic pressure set by $\bar{E} = 0.91 = \bar{P}_c^v$ by Eq.~\eqref{ratio}. 
		(b.) Eq.~\eqref{tc} collapses the same experimental data plotted in (a.), and captures the critical time for intermediate pressure values. When $\bar{P}$ decreases to about $0.94$, the theory underpredicts the critical time. Error bars correspond to one standard deviaton in $\bar{P}$. Horizontal dashed lines in a. \& b. indicate the elastic snap-through time $\tau_{inertial}$, which sets the minimum $t_c$.}
	\label{fig:tc}
\end{figure}

Eq.~\eqref{tc} is plotted against our data in Fig.~\ref{fig:tc}. Like for the critical deflection (Fig.~\ref{fig:wc}), the knockdown theory captures the critical time well in the intermediate range. At or above the elastic limit $\bar{P} = 1$, Eq.~\eqref{tc} predicts nonphysical critical times of $t_c \leq 0$. This is because the SLS model assumes that both loading and initial elastic deformation happens instantaneously, so creep begins at $t=0$ s. Of course, in reality elastic instability occurs on a timescale associated with the elastic wavespeed. Accordingly, the inertial timescale $t_*$ begins to dominate the viscoelastic one as $\bar{P}$ approaches $1$. Following e.g. Refs.~\cite{Pandey2014,Gomez2016,Sieber2019}, we expect that the elastic timescale $t_* \sim (2 R)^2/(c h)$, where $c = \sqrt{E_0/\rho} = 23.37$ m/s is the speed of sound within the material and $\rho = 1080$ kg/m$^3$ is the material density according to the manufacturer. For our shells in order of increasing thickness, this gives $t_* \sim \{0.1027, 0.0493, 0.0264, 0.0173\}$ s. We have indicated the elastic snap-through time for an arch $\tau_{inertial} = 2 \sqrt 3 t_*$~\cite{Pandey2014}, with horizontal dashed lines in Fig.~\ref{fig:tc}a.

When the imposed pressure nears the lower limit $\bar{P}_c^v$, the asymptotic behavior is captured qualitatively (see Fig.~\ref{fig:tc}a). However, the theory underestimates the critical time in this region. This divergence occurs around $\bar{P} \approx 0.94$, which corresponds to when the predicted buckling time surpasses $\tau_{\sigma}$. Thus, deformation has slowed considerably prior to buckling for these experiments and the inertia that was present at early times is no longer available. We have identified elastic buckling in our experiments as when the deformation rate exceeds that at very early times (when the shell also behaves elastically). However, it is possible that buckling initiated sooner in reality, but that the shell needs further perturbation -- that is, to deform more, which requires more time -- before we detect collapse. This is reminiscient of critical slowing down phenomena, wherein dynamics slow considerably near instability~\cite{Gomez2016,Gomez2018}. Other possible explanations for the deviation from our model are the simplified representations of stresses and strains, or the inability of the SLS model to capture the material behavior exactly. Nonetheless, we conclude that despite the notable simplicity of our assumptions, the knockdown theory explains our observations quite well, and does so while emphasizing the close connections between elastic and viscoelastic shell buckling.

\section{Conclusion}\label{sect:concl}
In summary, we subjected thick, spherical, defect-seeded viscoelastic shells to step pressure loading. We observed three regimes: When the pressure load was at or above the experimentally determined elastic critical load, we observed predictible elastic behavior, i.e. prompt buckling. At intermediate loads -- below the elastic critical pressure -- the shell buckled, albeit after a time delay during which deformation slowly progressed. At still-lower pressures, the shell deformed but collapse never occured. Our aim in this work was to rationalize our findings in a way that maintains close ties to elastic shell buckling, and is readily useable for experiment or design goals. To this end, we demonstrated that the load thresholds, critical deflection, and critical time may all be captured by a framework that treats creep deformation like an evolving defect in an elastic shell.

In particular, the ratio of the long-term modulus to the short-term (elastic) one is the same as as the ratio of the two critical pressures. This result is rooted in elastic shell theory, but practically, the material properties alone can explain the two pressure thresholds. This finding was suggested in various theoretical works on creep buckling~\cite{Hayman1978,Hayman1981,Hoff1973} and is independent of geometry, and hence is completely general.

We used this fact and existing work~\cite{Lee2016} on defects in spherical shells to discern an expression for how deformation due to creep acts to ``knock down" the critical pressure. In this view, the shell loses stability when creep deformation, which localizes at pole and amplifies the initial imperfection, progresses enough to reduce the critical pressure to the value of the imposed one. This allowed us to capture the dependence of the critical deflection on the imposed pressure (normalized by the experimentally-determined elastic critical pressure), the modulus ratio, and the defect geometry (Eq.~\eqref{wc}). This offers an explanation rooted in elastic shell behavior for a decidedly viscoelastic phenomenon: in the delayed buckling regime, higher deflection is required for instability as the pressure decreases. 

Because deformation occurs on a timescale sufficiently well-described by our chosen viscoelastic material model (SLS), a time-dependent form of the viscoelastic knockdown function immediately follows. From this we devise an expression for how the pre-buckling delay time depends on the same quantities: the modulus ratio, the dimensionless pressure, and the defect geometry (Eq.~\eqref{tc}). The buckling time increases monotonically but non-linearly as the pressure decreases, which is generally captured by our model. While the modulus ratio $\bar{E}$ was fixed in our experiments, we expect that our model is valid for any material with relatively low relaxation strength (i.e. the relaxation modulus and creep compliance functions do not differ significantly). Further, we note that the success of our model does not depend on the approximation $J^{-1}(t) \approx E(t)$ that we have employed. Rather, if one were to obtain the creep compliance function in experiments, using this directly in place of the relaxation modulus would likely improve the accuracy of predictions.

While viscoelastic shells behave elastically during buckling, viscous effects re-enter in later stages of postbuckling, as discussed in Ref.~\cite{Coupier2019}. Unbuckling was studied in detail in Ref. \cite{Djellouli2017}, as the non-reciprocal nature of the buckling-unbuckling cycle is the source of motility. We did not examine unbuckling in the present work. However, we could reasonably expect that another viscoelastic delay phenomenon, termed ``pseudo-bistability"~\cite{Brinkmeyer2012} (first introduced in Ref.~\cite{Santer2010} as ``temporary bistability"), could be seen in our shells. 

Pseudo-bistability refers to the delayed ``snap-back" instability that occurs in the unloaded state, following a loading-unloading sequence that induces both stress-relaxation and creep~\cite{Gomez2018}. Early works modeled viscoelasticity \textit{via} an evolving stiffness, acheiving qualitative agreement with experiments~\cite{Santer2010,Brinkmeyer2012}. A recently-proposed metric framework introduces viscoelasticity as a temporally evolving (fictitious) reference length instead, and among its merits is the ability to predict delayed snap-back instability~\cite{Urbach2020,Urbach2018}. Delayed snap-back instability occurs in the unloaded state, and in our setting would require maintaining the pressure load for a sufficiently long time while the shell is fully buckled before unloading. As such, creep instability and pseudo-bistable snap-back can in theory be induced in the same thick structure, with one or the other surpressed at will. 

Our findings highlight the importance of the load rate, and the sensitivity of the shell to pressure variations in the vicinity of the elastic critical load (when approached from below). These are important considerations for structural designs using silicone rubber where either stability or elastic behavior is desired. For instance, knowledge of the critical load thresholds is clearly important when designing an efficient spherical swimmer, or indeed any viscoelastic structure that should undergo oscillatory instability. If the goal is fast motility, it would likely be desirable to minimize the time delay before buckling by avoiding the delayed buckling (intermediate pressure) regime altogether. 

In other settings, viscoelastic behavior can enhance the functionality of reversibly actuatable structures. This has been demonstrated recently in designs that rely on pseudo-bistable snap-back, \textit{e.g.} 3D-printed viscoelastic metastructures whose time-dependent properties are tunable based on temperature~\cite{Che2019}, with even more flexibility afforded by using multiple viscoelastic materials~\cite{Che2018}. Another study examines the interplay between viscous dissipation and geometric hysteresis, as a function of the strain rate, for the design of optimal energy dissipating metamaterials~\cite{Dykstra2019}. Introducing tunable delays \textit{via} creep buckling has not yet been explored, but the potential is vast: A switch or capsule could be loaded subcritically, so that the shell has time to move to a desired location before buckling occurs. Over time as deformation reduces the critical load and thus the energy barrier~\cite{Hutchinson2018}, a much smaller probing force or other perturbation could trigger buckling. This could be useful for pneumatic gripping. A mechanical signal of fixed input frequency (and varying amplitude) could produce a varied output frequency, which has implications for mechanical computing. Because our analysis relies only on quantities that are straightforward to determine in experiments, our findings are especially amenable to accessing such tunability. As we have shown, these possibilities are achievable with the nearly-elastic materials that are already common and cherished in mechanics research.

In 1956, Gerard wrote about creep buckling that ``there are almost more theories than reliable test points which can be used to check the theories"~\cite{Gerard1956}. While a limited number of experimental contributions have come about since, to our knowledge our experiments on full spheres are novel to the existing creep buckling literature. Further, we have demonstrated that some concepts central to general creep buckling theories, which previously were mostly tested on metallic, mono-resin materials, and reinforced concrete~\cite{Miyazaki2015}, are indeed applicable to soft, rubbery materials. We expect that the concepts we have studied can be extended to other geometries, other loading methods, and similar polymers.

\section*{Conflicts of interest} \noindent We declare that we have no competing interests.

\section*{Funding} \noindent D.P.H. and L.S.M. are grateful for financial support from the NSF CMMI-1824882. G.C. acknowledges financial support from the European Community's Seventh Framework Programme (FP7/2007-2013) ERC Grant Agreement Bubbleboost no. 614655.

\section*{Acknowledgements} \noindent We thank Catherine Quilliet and Philippe Marmottant for many helpful discussions, Adel Djellouli for the shell fabrication protocols and equipment, Patrice Ballet for his technical support, and Francisco Lopez-Jimenez and Pedro Reis for providing data from their work in Ref.~\cite{Lee2016}
	

\section{Appendix}

\subsection{SLS parameters}\label{sect:creeptests}
To determine the material properties of Dragon Skin 30\textsuperscript{TM}, we fabricated 7 dogbone specimens (ASTM D412 Type C). The molds, cut using the Epilog Helix laser machine, were designed to give each sample a protruding defect -- two thin horizontal lines 500 microns thick and separated by 500 microns -- at the middle of the gauge section. These lines were tracked with a zoom lens attached to a Nikon D610, allowing for accurate strain measurements taken later in ImageJ. Samples were then tested after 16, 18, 24, 44, 65, and 94 hours at room temperature post-fabrication, as well as 1 hour after 25 minutes of curing at an elevated temperature of 65$\degree$C. The manufacturer lists the cure time as 16 hours.

\begin{figure}[h!] 
	\centering
	\includegraphics[width=0.7\linewidth]{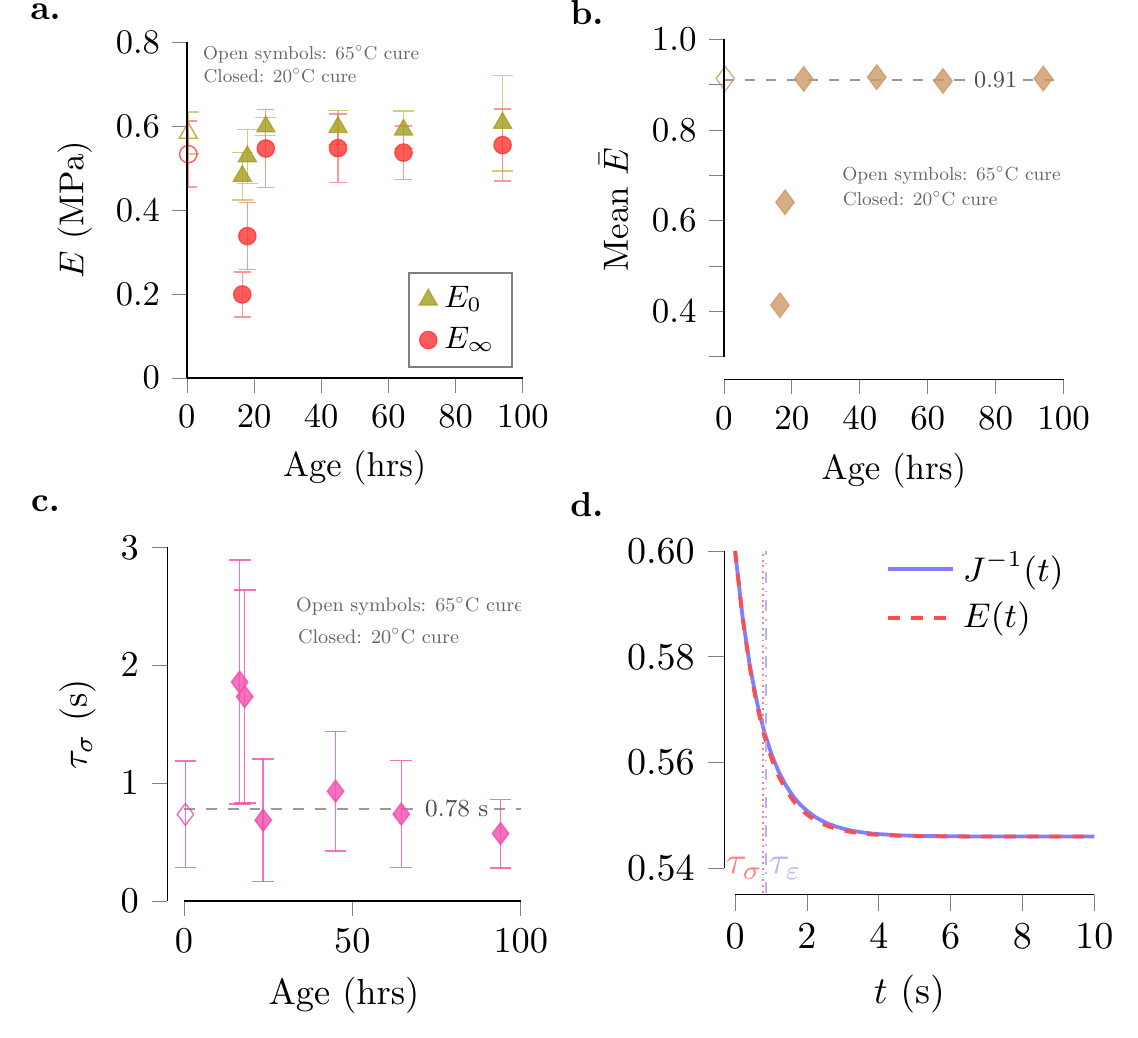}	
	\caption{Averaged values for (a.) the elastic and long-term moduli of Dragon Skin 30\textsuperscript{TM} and (b.) the ratio of their mean values, as determined from tensile stress-relaxation tests. The x-axes indicate the time elapsed since fabrication; The cure time is listed by the manufacturer as 16 hours post-fabrication. (c.) Plot of the relaxation time from stress-relaxation tests. The dashed lines in (b.) \& (c.) show the averaged values used in our analysis, $\bar{E} = E_{\infty}/E_0 = 0.91$ and $\tau_{\sigma} = 0.78$ s, which were determined from the samples whose ages were relevant to our experiments (24-48 hours cured at room temperature, and 0.4 hours cured at elevated temperature). (d.) SLS relaxation modulus ($E(t)$) and inverse creep compliance ($J^{-1}(t)$) functions constructed from our averaged quantitites. Because the material is nearly elastic, the $E(t) \approx J^{-1}(t) \: \forall \: t>0$. All error bars represent one standard deviation.}
	\label{fig:creepparams}
\end{figure}

The long-term (equilibrium) modulus was determined from tensile stress-relaxation tests using the tensile testing machine Instron 5943. Displacements, which resulted in strains $\varepsilon_0 \in [2\%,20\%]$, were imposed at rates of 150-500 mm/min, then maintained for 120 seconds while the force, which decreases over a timescale $t \sim \tau_{\sigma}$ before plateauing, was measured. The plateau force was used to calculate $\sigma_{\infty}$ where we took $E_{\infty} \equiv \sigma_{\infty}/\varepsilon_0$. The values for $E_{\infty}$ plotted in Fig.~\ref{fig:creepparams}a are calculated as the average of 12 total measurements from 4 tests at varied strain levels and strain rates. 

This relatively fast loading resulted in measurement uncertainty at early times, so we did not fit a curve to the entire range of stress-relaxation data, nor did we extract the elastic modulus from stress-relaxation tests. Instead, tensile tests on the same samples were conducted at rates of 10 and 20 mm/min up to $\varepsilon_{\text{max}} \in [8\%,26\%]$. This resulted in linear stress-strain curves, and the slopes were used to calculate $E_0$. Each data point in Fig.~\ref{fig:creepparams}a represents the average of 6 measurements from 2 tests for each sample. 

Between 16 and 24 hours post-fabrication at room temperature (approximately 20\degree C), we observe an increase in the long-term modulus from $0.20 \pm 0.05$ MPa to $0.55 \pm 0.09$ MPa. By 24 hours, the long-term modulus reaches a plateau. The elastic modulus follows a similar trend: the material stiffens from $E_0 = 0.48 \pm 0.06$ MPa at 16 hours post-fabrication to $0.60 \pm 0.02$ MPa at 24 hours, by which time the elastic modulus has plateaued. The plateau value we measure is in agreement with the 100\% modulus value listed by the manufacturer, Smooth-On, of 0.59 MPa. More or less the same plateau values result from curing the sample in the oven at 65\degree C for 25 minutes ($E_{\infty}=0.53 \pm 0.08$ MPa, $E_0 = 0.58 \pm 0.05$ MPa). 

The ratio $\bar{E} = E_{\infty}/E_0$ is central to our analysis, and is plotted in Fig.~\ref{fig:creepparams}b. For the shell-buckling experiments discussed in the body of this paper, the relevant times are 24 and 44 hours, and the 25 minute oven-cured sample (see Sect. \ref{sect:shellfab}). Averaging this data gives the value we use throughout our analysis (and the dashed line in Fig.~\ref{fig:creepparams}b), $\bar{E}  = 0.91$. Instead including all of the samples whose moduli appear to have saturated (i.e. all except the 16 and 18 hour samples) does not change this averaged value (it only slightly increases the standard deviation, to 0.3 MPa.)  

With the functional form of the relaxation modulus in mind, we identified the relaxation time $\tau_{\sigma}$ as the time when the modulus has reduced to $E_{\infty}+E_1 e^{-1}$. The averaged values are shown in Fig.~\ref{fig:creepparams}c. The overall average was $\tau_{\sigma} = 0.78 \pm 0.49$ s. This corresponds to a retardation time $\tau_{\varepsilon} \approx 0.86$ s~\cite{Lakes2009}. The relatively large error range on the characteristic timescale results from initial uncertainty in stress-relaxation curves.

In Sects. \ref{sect:press} \& \ref{sect:defect}, we blur the lines between the creep compliance and the relaxation modulus: While the active process is creep, we use not the creep compliance, but the relaxation modulus -- which describes stress-relaxation -- in our analysis. This is for two reasons: First, our primary aim is to describe creep buckling in the language of elastic buckling -- which refers to Young's Modulus $E$ -- rather than to provide an exact description of creep in this material. Second, we found that stress-relaxation (displacement-controlled) tests provided much more reliable data than creep (force-controlled) tests using Instron 5943.

The relaxation modulus (Eq. \eqref{relaxmod} in the main text) is related by Laplace transform to the creep compliance (Eq. \eqref{creepcomp}). Thus, the limiting values $E_0 = \sigma(t=0)/\varepsilon$ and $J_0 = \varepsilon(t=0)/\sigma$; $E_{\infty}= \sigma(t=\infty)/\varepsilon$ and $J_{\infty} = \varepsilon(t=\infty)/\sigma$ are exact inverses. 

The relaxation time $\tau_{\sigma}$ is less than the retardation time $\tau_{\varepsilon}$. The two are related through the relaxation strength, defined as $\Delta = E_1/E_{\infty}$, according to $\tau_{\varepsilon} = \tau_{\sigma} (1+\Delta)$, and thus $E(t) \neq J^{-1}(t) \forall t$~\cite{Lakes2009}. However, the difference between the two functions is very small, as shown in Fig.~\ref{fig:creepparams}d. Due to this negligible difference, we conclude that our approximation in Sect. \ref{sect:defect} is also reasonable.

\subsection{Shell fabrication}\label{sect:shellfab}
\begin{figure}[h!] 
	\centering
	\includegraphics[width=.73\linewidth]{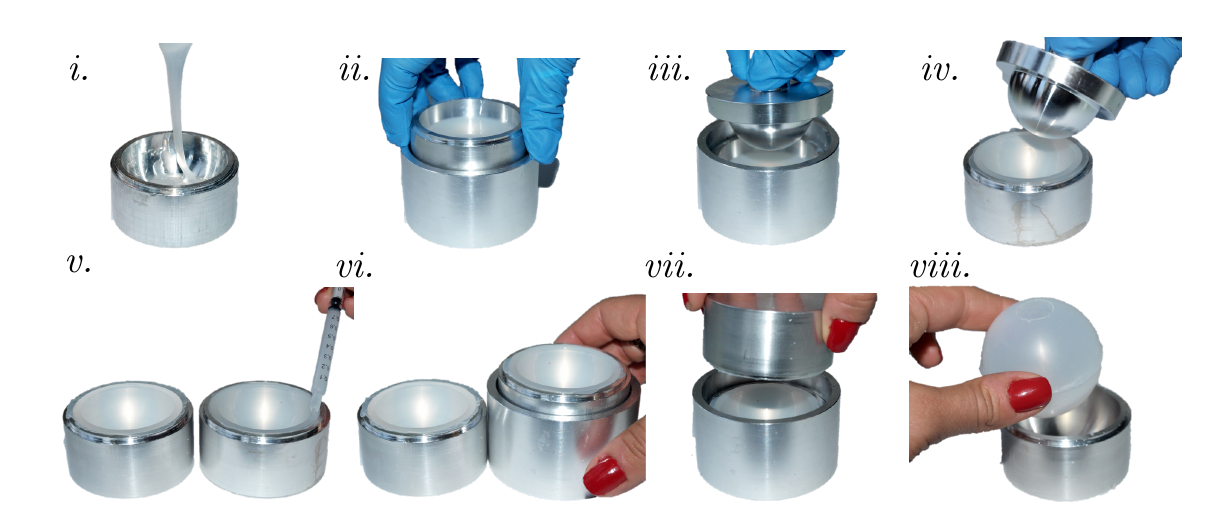}
	\caption{Shell fabrication process. \textit{i}. Mixed and degassed DragonSkin 30 is poured into two hemispherical aluminum cavities of $R_o = 25$ mm. Only one is shown in \textit{i.}-\textit{iv.}, and several layers of tape have been cut into a circle and adhered to the center of one of the hollow spheres. \textit{ii.} After degassing again, the filled cavity is fitted with an alignment sleeve. \textit{iii.} A half-sphere of $R_i < R_o$ is inserted. \textit{iv.} After curing for 25 minutes in the oven, a spherical hemisphere has formed. \textit{v.} Heptane-diluted polymer is deposited with a syringe to glue two cured hemispheres. \textit{vi.-vii.} The edges of the two hemispheres are joined and aligned by the alignment sleeve. \textit{viii.} After curing at room temperature for at least $16$ hours, a sealed sphere is removed from the mold.}
	\label{fig:shellfab}
\end{figure}
The process for fabricating polymeric spherical shells, which was developed in \cite{Djellouli2017}, is as follows: Custom aluminum molds consist of female and male components along with alignment sleeves. The female mold is a cylinder of equal radius and height ($30$ mm) with a hemispherical cavity, which sets the outer radius of the shells to $R_o=25$ mm. The male component consists of a shouldered half-sphere whose size determines the inner shell radius, $R_i=\{24, 23, 21.5, 20\}$ mm. The shoulder (chamfered to $5\degree$) is $10$ mm tall, and its maximum diameter matches that of the female mold as well as the inside of the guiding sleeve. The latter is a $40$ mm tall cylindrical tube, internally chamfered up to a depth of $10$mm to accommodate the male mold. 

To control the location of the onset of buckling, shells are seeded with a circular imperfection. This is achieved by affixing 1-4 layers of adhesive tape cut to $R_{\delta} \approx 6$ mm (resulting in imperfection depth of $\delta \in \{0.76, 0.81, 0.84, 0.40\}$ mm, in order for the thinnest to thickest shell) to the center of one of the two female molds used to make each shell.

To make shells, the polymer is prepared according to package instructions, degassed in a vacuum, and poured into the female molds. After degassing again, each polymer-filled cavity is fitted inside an alignment sleeve, and the male mold is inserted. The assembly is tightly clamped between the two plates of a simple mechanical press and cured at 65\degree C for 25 minutes. 

After curing, the alignment sleeve and the male component are removed, revealing two hemispherical shells resting in the female molds. To join the two halves, a glue is prepared. The viscosity of the liquid polymer is reduced via dilution with heptane at a 2:1 ratio. This allows for the application of a sufficiently thin layer of glue, deposited with a syringe around the equator of each hemisphere. Again a sleeve is used to align the two halves, whose contact is ensured by the mechanical press, and the shell is left to cure at room temperature for 16 hours. 

A drill press is used to create a 1 mm diameter hole, into which a small nozzle connected to a tube allows for internal pressure control. Lastly, a suction cup ($2-3$ cm diameter) is glued to the shell surface opposite the buckling spot with cyanoacrylate (Loctite). During experiments, the shell is fixed in place via a screw attached to the back of the suction cup. 

\subsection{Knockdown of elastic critical pressure due to through-thickness defects}\label{sect:PceSect}
Unsurprisingly, the classical prediction for the buckling pressure of a perfect elastic spherical shell (Eq.~\eqref{ZoellyPc}) does not capture the behavior of our imperfect shells. Recently, predictions for the knockdown factor as a function of the size of an axisymmetric imperfection have been presented for dimple-like~\cite{Lee2016,Jimenez2017} and through-thickness~\cite{Yan2020} defects.
\begin{figure}[h!] 
	\centering
	\includegraphics[width=0.4\linewidth]{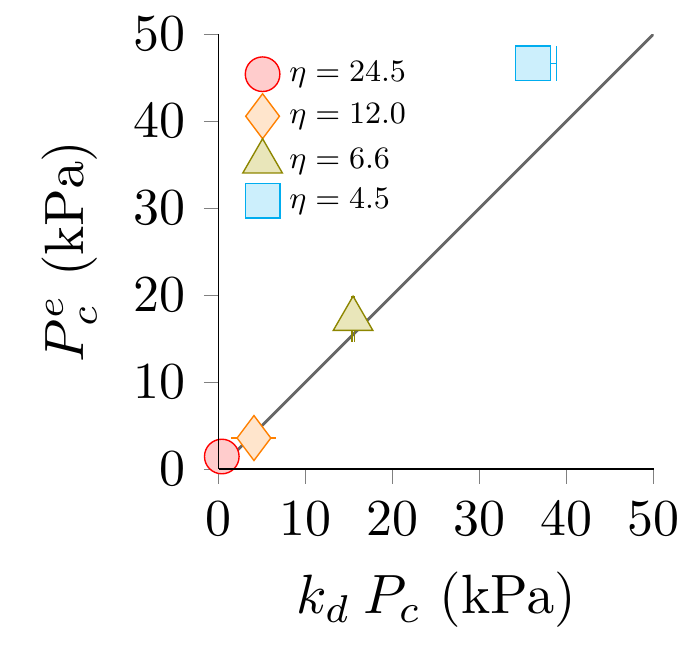}
	\caption{Comparison of the theoretical elastic buckling pressure $P_c^e$ for thin, shells with through-thickness imperfections with our experimental value. The experimental value is taken as the minimum pressure at which no slowdown of deformation occurs before buckling. The theoretical value $k_d P_c$ is calculated using $k_d$ extracted from Fig. 11b in Ref.\cite{Yan2020}, and $P_c$ from Eq~ \eqref{ZoellyPc}. Error bars correspond to approximately one standard deviation, and are smaller than the markers in most cases.}
	\label{fig:pcefig}
\end{figure}

The reduced-thickness defects in our experiments (see Fig.~\ref{fig:shellparams}) are like those in the work of Yan et al.~\cite{Yan2020}. The authors present data from experiments and FEM simulations on the knockdown factor for varied depth, angular width, and transition width of the imperfection. We use this data\footnote{Knockdown values were extracted from Figure 11b of Ref.~\cite{Yan2020}, and found to be $k_d = \{0.2719 \pm 0.007, 0.7423 \pm 0.000, 0.8664 \pm 0.008, 0.93 + 0.070\}$ for $\eta = \{24.5,12.0,6.4,4.5\}$, respectively.} to calculate the ``knocked-down" theoretical elastic critical pressure for each of our shells. This theoretical value is plotted against our experimental data in Fig. \ref{fig:pcefig}. We have taken $P_c^e$ to be the minimum pressure at which a plot of volumetric or pole deformation versus time shows no slope change until buckling (the rate of deformation is much slower when creep occurs.)

Up to relatively thick shells, the theoretical knockdown predicts our experimentally measured critical pressure relatively well ($\eta = 24.5,12.0,6.6$) -- perhaps surprsingly so, given that the data in Ref.~\cite{Yan2020} was collected in quasi-static experiments on shells of fixed $\eta = 100$. However, as $\eta$ decreases even further from the thin shell limit, the thick shell withstands higher pressures than predicted. This inapplicability of thin shell knockdown theories is significant for the thickest shell we tested ($\eta = 4.5$). For consistency, then, in all arguments throughout the main text we rely on an experimentally determined value for the elastic buckling pressure. Thus, we take the experimental value for $P_c^e$ to be the minimum pressure difference where elastic buckling occurs.

\subsection{Methods for dynamic pressure loading experiments}\label{sect:pressurecontrol} 
Before each experiment, a vacuum pump (Becker U 4.40) is used to reduce the pressure inside a 50 liter tank. A flexible tube ($3$ mm inner diameter, $12$ cm length) connects the tank to a differential pressure sensor (Freescale Semiconductor MPX5100DP, sensitivity 45mV/kPa). Pressure readings were recorded using a microcontroller (Arduino UNO) every $0.02$ s with a resolution of $0.1$ kPa. 

As this pressure resolution is slightly coarse for the thinner shells, which buckle at pressures on the order of $1$ kPa, we account for rounding (and small pressure fluctuations) by reporting the mean and standard deviation of the pressure values reported at intermediate\footnote{For shells subjected to large $P$ which buckle elastically, we simply report the (non-fluctuating) imposed pressure, recorded before the shell begins to deform, with a default maximum error of $\pm 0.05$ kPa.} times (after initial elastic deformation and before buckling, should it occur).

Via a T-junction, a second tube ($6$ cm long) connects the tank to the inside of the shell, by way of an electrovalve (Matrix Pneumatics Solenoid Valve, MX891.901C224). The response time of the valve is $<1$ ms. Based on a back-of-the-envelope calculation of the amount of gas that needs to travel from the shell to the tank for pressure equilibrium and the corresponding mass flow rate, we expect that shell-vacuum tank system equilibriates within approximately $4$ ms for pressures imposed on the thickest shell (where higher pressure gradients drive faster air flow), to about $40$ ms for the thinnest.

Arduino IDE software enables synchronization of the sudden opening of the valve with the digital recording of the pressure difference (at a rate of 500 Hz), as well as with the triggering of the high-speed camera (Phantom Miro 310). Depending on the empirically determined delay time before buckling, images were captured at a rate of 2000-9000 frames per second. Image processing of TIFF stacks was completed using ImageJ and custom Python scripts.


\end{document}